\documentclass[sigconf]{acmart}
\usepackage{graphicx}
\usepackage{hyperref}
\usepackage{todonotes}
\usepackage{pifont}
\usepackage{subcaption}
\usepackage{array}
\usepackage{multirow}
\usepackage{enumitem}

\acmConference[GeoAnomalies'24]{Make sure to enter the correct
   conference title from your rights confirmation email}{October 29, 2024}{Atlanta, GA, USA}

\def\refsec#1{Section~\ref{#1}}

\def\reftab#1{Table~\ref{#1}}

\def\GitHubRepo{\url{https://github.com/onspatial/sigspatial2024-anomaly-dataset}}

\newcommand*{\crossmark}{$\times$}
\newcommand{\rulesep}{\unskip\ \vrule\ }

\renewcommand\footnotetextcopyrightpermission[1]{}

\begin{document}

\title{Urban Anomalies: A Simulated Human Mobility Dataset with Injected Anomalies}
\author{Hossein Amiri}
\email{hossein.amiri@emory.edu}
\affiliation{%
    \institution{Emory University}
    \city{Atlanta}
    \country{USA}
}

\author{Ruochen Kong}
\email{ruochen.kong@emory.edu}
\affiliation{%
    \institution{Emory University}
    \city{Atlanta}
    \country{USA}
}

\author{Andreas Z{\"u}fle}
\email{azufle@emory.edu}
\affiliation{%
    \institution{Emory University}
    \city{Atlanta}
    \country{USA}
}

\begin{abstract}
    
Human mobility anomaly detection based on location is essential in areas such as public health, safety, welfare, and urban planning. Developing models and approaches for location-based anomaly detection requires a comprehensive dataset. However, privacy concerns and the absence of ground truth hinder the availability of publicly available datasets. 
With this paper, we provide extensive simulated human mobility datasets featuring various anomaly types created using an existing Urban Patterns of Life Simulation.
To create these datasets, we inject changes in the logic of individual agents to change their behavior. Specifically, we create four of anomalous agent behavior by \
(1) changing the agents' appetite (causing agents to have meals more frequently), 
(2) changing their group of interest (causing agents to interact with different agents from another group).
(3) changing their social place selection (causing agents to visit different recreational places) 
and 
(4) changing their work schedule (causing agents to skip work), 
For each type of anomaly, we use three degrees of behavioral change to tune the difficulty of detecting the anomalous agents.
To select agents to inject anomalous behavior into, we employ three methods: (1) Random selection using a centralized manipulation mechanism, (2) Spread based selection using an infectious disease model, and (3) through exposure of agents to a specific location.
All datasets are split into normal and anomalous phases. The normal phase, which can be used for training models of normalcy, exhibits no anomalous behavior. The anomalous phase, which can be used for testing for anomalous detection algorithm, includes ground truth labels that indicate, for each five-minute simulation step, which agents are anomalous at that time. Datasets are generated using the maps (roads and buildings) for Atlanta and Berlin having 1k agents in each simulation. All datasets are openly available at \url{https://osf.io/dg6t3/}. Additionally, we provide instructions to regenerate the data for other locations and numbers of agents.

\end{abstract}
\renewcommand{\shortauthors}{Amiri, et al.}

\maketitle


\section{Introduction}

Human mobility data holds vital information that can be harnessed to gain valuable insights across various applications~\cite{lv2017outlier}. A critical aspect of working with GPS data is anomaly detection, particularly location-based anomaly detection, which plays a key role in numerous algorithms designed for both industrial and research purposes~\cite{smiti2020critical, belhadi2020trajectory, kriegel2010outlier}. One of the major challenges in developing models and algorithms for anomaly detection is the scarcity of ground-truth information in publicly available real-world location data~\cite{zhang2023large}. Anomaly activities are often underreported or missing in real-world datasets, making it difficult to validate and refine detection methods. Consequently, many researchers turn to synthetic data to develop their models, whether in supervised or unsupervised frameworks~\cite{lohrer2023gadformer}.

Anomaly detection algorithms have been extensively studied in the literature \cite{lee2008trajectory,yu2017outlier,zhang2023large,lv2017outlier}. However, the absence of a standardized benchmark dataset remains a critical challenge in this domain \cite{campos2016evaluation,stanford2024numosim}. Researchers often rely on various real-world datasets to evaluate their algorithms, but these datasets typically lack ground truth information. The UCI GPS database comprises 603 car driving trajectories with over 5,317 points, where each trajectory contains more than 2,000 points, making it relatively sparse \cite{cruz2015grouping}. Similarly, the Geolife trajectory database, collected by Microsoft Research Asia, includes 17,621 trajectories with over 152,241 points, with each trajectory exceeding 5,000 points, also contributing to sparsity \cite{zheng2011geolife}. In contrast, the T-Drive trajectory database, also from Microsoft Research Asia, represents a dense dataset generated by over 33,000 taxis in a period of 3 months which does not have ground truth for anomaly detection tasks~\cite{yuan2010t}. Furthermore, the field would greatly benefit from the development of large-scale, simulated datasets that incorporate human mobility, check-ins, and social networks, such as those generated using location-based social network simulation models \cite{amiri2023massive}.

In this paper, we leverage an existing Pattern of Life Simulation model~\cite{zufle2023urban,kim2020location} to simulate and identify anomaly behaviors in agent-based environments. This simulation roots the behavior of agents on Maslowian Needs~\cite{maslow1943theory}, such as Physiological Needs (e.g. Food Need), Safety Needs (e.g. Financial Need), and Love Needs (e.g. a need to maintain friendships). Details on these needs and their prioritization that leads to agent actions and behavior are detailed in~\cite{zufle2023urban}. 
To add anomalous behavior into this simulation, we define anomalies as deviations from normal patterns that suggest they were generated by a different underlying mechanism~\cite{hawkins1980identification,barnett1994outliers}. We inject four types of anomalies at three different intensity levels to simulate abnormal patterns of life, using three distinct mechanisms to inject these anomalies.
%
To model anomalous agent behavior, we inject four types of anomalous behavior: 
\begin{itemize}[noitemsep,nosep,leftmargin=20pt,labelsep=5pt,itemindent=0pt]
    \item {\bf{Hunger Anomalies}}, which increases the rate at which the ``Food Need'' of agents increases over time. Hunger anomalies lead to anomalous patterns of life due to requiring agents to visit places that offer food more frequently, thus disrupting the agents'     normal patterns of life,
    \item {\bf{Interest Anomalies}}, which changes agents' interest attribute, which is a nominal attribute that defines which type of recreational agents prefer to visit. 
    \item {\bf{Social Anomalies}}, which cause agents to replace their normal choice of recreational sites, which is based on their interests and their social connections with other agents they often meet at these sites, which a random choice of recreational sites. This behavior disrupts the agents' social networks due to preventing them from meeting their friends and meeting random agents instead.
    \item {\bf{Work Anomalies}}, which cause agents to stop going to work, giving them more time to focus on non-work related needs and activities,
\end{itemize}
For each type of anomaly, we define three levels of intensity of anomalous behavior: Yellow (low intensity), Orange (moderate intensity), and Red (high intensity).

To choose which agents to select as anomalous and when to cause them to follow their anomalous behavior, we employ three different infection mechanisms: centralized injection, infectious disease spread, and location-based, each tailored to address different research questions. 
\begin{enumerate}[noitemsep,nosep,leftmargin=20pt,labelsep=5pt,itemindent=0pt]
    \item A {\bf{Centralized Model}} injects anomalies randomly through a centralized manipulation mechanism. All randomly selected agents adopt anomalous behavior during a specified (``Test'') period. 
    \item An {\bf{Infectious Disease-Based Model}}, which employs a Susceptible-Infectious-Recovered (SIR) infectious disease model~\cite{kohn2023epipol}. Only a relatively small number of agents is selected randomly using the centralized approach as initial infections. Infected agents may then infect other agents through co-location at the same place of interest. After a specified duration, agents recover, are no longer infectious, and cannot be re-infected.
    \item A {\bf{Location-Based Model}}, which selects a place of interest as the source of anomalous behavior. Any agent visiting this place has a chance to become anomalous. This model simulates scenarios like the 1854 outbreak of cholera in London that was caused by a particular public water pump discovered by John Snow's spatial analysis of cholera cases~\cite{snow1855mode}. 
\end{enumerate}

For each of the aforementioned types of anomalous behaviors, and for each type of injection mechanism, we provide data resulting from a simulation of 1000 agents, each having four weeks of normal behavior (without any anomalies) and having at least four weeks of anomalous behavior (during which anomalies are injected as described above).
We hope that these datasets will allow scholars to investigate novel approaches to identify anomalous behavior in simulated human mobility data.

\begin{table}[t]
    \centering
    \caption{ Simulated Realistic Human Mobility Features: A checkmark (\checkmark) indicates that the dataset provided in this paper includes useful information. A crossmark (\crossmark) indicates that the data is either unavailable or consists of static values that do not provide meaningful insights. \vspace{-0.3cm}}
    \label{tab:information_clarification}
    \begin{tabular}{|c|c|c|}
        \hline
        \multirow{7}{*}{\centering\rotatebox{90}{\textbf{Mobility}}}    & \textbf{Information}  & \textbf{Urban Anomalies} \\ \cline{2-3}
                                                                        & GPS Trajectories      & \checkmark               \\ \cline{2-3}
                                                                        & Staypoints            & \checkmark               \\ \cline{2-3}
                                                                        & Check-Ins             & \checkmark               \\ \cline{2-3}
                                                                        & Stay Duration         & \checkmark               \\ \cline{2-3}
                                                                        & Trip Duration         & \checkmark               \\ \cline{2-3}
                                                                        & Trip Purpose          & \checkmark               \\ \cline{2-3}
                                                                        & POI Semantics         & \checkmark               \\ \hline
        \multirow{5}{*}{\centering\rotatebox{90}{\textbf{Kinematic}}}   & Velocity              & \crossmark               \\ \cline{2-3}
                                                                        & Acceleration          & \crossmark               \\ \cline{2-3}
                                                                        & Traffic Lights        & \crossmark               \\ \cline{2-3}
                                                                        & Traffic Congestion    & \crossmark               \\ \cline{2-3}
                                                                        & Walking/Gait Patterns & \crossmark               \\ \hline
        \multirow{4}{*}{\centering\rotatebox{90}{\textbf{Social}}}      & Friendship            & \checkmark               \\ \cline{2-3}
                                                                        & Social Interactions   & \checkmark               \\ \cline{2-3}
                                                                        & Social Network        & \checkmark               \\ \cline{2-3}
                                                                        & Communities           & \checkmark               \\ \hline 

    \end{tabular}
    \vspace{-0.65cm}
\end{table}

We note that Urban Anomalies data does not include kinematic features; while kinematic features like speed and acceleration vary in the real world due to traffic conditions, traffic lights, and STOP signs, Urban Anomalies assumes a constant-speed movement on the shortest path. Therefore, Urban Anomalies datasets simulate data on human mobility (where do people go and why?) but not on human kinematics (how do they get there?). For more details, \reftab{tab:information_clarification} shows the types of human mobility features, which are realistically simulated in Urban Anomalies.
With these insights, our main contributions are as follows:

\begin{itemize}[noitemsep,nosep,leftmargin=20pt,labelsep=5pt,itemindent=0pt]
    \item We customize an existing Pattern of Life Simulation model to simulate and identify anomaly behaviors in agent-based environments.
    \item We provide trajectory, stay points, and social links for 100+ different settings that are split into normal life (train) and normal life an anomaly life (test)
    \item We inject three distinct approaches to categorize the start of their anomalies' life, such as centralized manipulation, infectious diseases model, and location-based infectious model
    \item Each dataset incorporates different types of anomalies and varying intensity levels and publicly available on~\url{https://osf.io/dg6t3/}.
    \item We conduct a comprehensive statistical analysis and visualize the anomaly behaviors in the simulated data.
\end{itemize}

The remainder of this paper is organized as follows.
\refsec{sec:related-work} reviews related works in the field of trajectory datasets and anomaly detection and discusses the data they used.
\refsec{sec:methodology} describes the data generation methodology used in our study, including the simulated anomalies and data processing.
\refsec{sec:description} presents the dataset description, including specifications, and analysis of the generated data.
\refsec{sec:regeneration} discusses the dataset regeneration process and instructions for reproducing the study.
Finally, \refsec{sec:conclusion} concludes the paper with a summary of our contributions and future research directions.
\vspace{-0.2cm}
\section{Related Work}
\label{sec:related-work}
A critical aspect of working with GPS data is the detection of anomalies, especially location-based anomalies. This process is fundamental to various algorithms employed in both industrial applications and research studies. Accurate identification and management of such anomalies help ensure the reliability and precision of models that rely on spatial data, which can greatly affect the outcomes of analyses ranging from navigation systems to geographical research~\cite{kriegel2010outlier}. Anomaly detection algorithms have been the focus of extensive research in the literature~\cite{lee2008trajectory,yu2017outlier,zhang2023large,zhang2024transferable,lv2017outlier,liu2024neural}. Despite this, a significant challenge persists in the form of the absence of a standardized benchmark dataset for the field~\cite{campos2016evaluation}. The lack of such a benchmark complicates the comparison and evaluation of different algorithms, as researchers are often compelled to rely on a variety of real-world datasets that typically lack ground truth information. This absence of consistent, labeled data makes it difficult to objectively assess the performance and generalizability of anomaly detection methods across different contexts and applications.

One of the primary challenges in developing models and algorithms for anomaly detection is the limited availability of ground-truth data in publicly accessible real-world location datasets. This scarcity of verified, accurate data complicates the validation and fine-tuning of models, thereby diminishing the effectiveness and reliability of anomaly detection methods. Consequently, researchers often encounter significant obstacles in creating robust solutions capable of accurately identifying anomalies in diverse and complex spatial datasets~\cite{zhang2023large}. The underreporting or complete absence of anomaly activities in real-world datasets further exacerbates these challenges, making it difficult to validate and refine detection methods. This lack of comprehensive data impedes the development of accurate models, as the absence of true anomalies hinders the assessment of detection algorithms' effectiveness. To address this limitation, many researchers turn to synthetic data for model development and testing, whether operating within supervised or unsupervised frameworks. Synthetic datasets offer a controlled environment where anomalies can be precisely defined and manipulated, enabling more rigorous testing and optimization of anomaly detection techniques~\cite{lohrer2023gadformer}. The flexibility and customizability of synthetic data make it an invaluable resource for researchers seeking to develop robust and adaptable models for anomaly detection in spatial datasets.

Trajectory anomaly detection algorithms are crucial for identifying anomalous patterns in spatiotemporal data, with significant implications for applications such as traffic management, urban planning, and movement pattern analysis. Numerous algorithms have been proposed to tackle this challenge, including depth-based, deviation-based, distance-based, density-based, and high-dimensional approaches, each contributing valuable insights into global and local anomaly detection methods~\cite{kriegel2010outlier}. For instance, the Density-Based Anomaly Trajectory Detection (DBOTD) algorithm, proposed in~\cite{lv2017outlier}, stands out by utilizing the Density-Based Spatial Clustering of Applications with Noise (DBSCAN)~\cite{ester1996density} to cluster trajectories and identify core routes as representatives, thereby improving both the speed and accuracy of anomaly detection. This study demonstrates DBOTD's effectiveness using a real-world dataset containing 5,660,692 trajectories in Beijing, where approximately 5,300 trajectories were manually labeled by volunteers as anomalies or not, serving as the testing set. The remaining trajectories were used for training, while the Beijing road network data, comprising 253,180 vertices and 557,134 edges, provided a comprehensive framework for evaluation. The results highlight DBOTD's superiority in both accuracy and efficiency compared to existing methods.

A comprehensive survey on trajectory analysis within the video surveillance domain, with a particular emphasis on trajectory anomaly detection, is presented in \cite{belhadi2020trajectory}. This survey examines the extensive research conducted in various industrial applications, such as maritime, urban transportation, and networking. It offers an in-depth overview of existing trajectory anomaly detection algorithms, categorizing them according to different criteria, including application, output, and algorithmic levels. The paper also identifies the key challenges and open issues within the field. Furthermore, it provides an extensive list of publicly available datasets used for trajectory analysis, including notable examples like the UCI GPS, the Geolife trajectory, and the Manhattan taxi datasets. 

Real-world trajectory datasets are commonly used for knowledge discovery in human mobility research, providing valuable insights into urban planning, transportation, and public health.
The GeoLife GPS trajectory dataset \cite{zheng2011geolife}, collected by Microsoft Research Asia, is widely recognized for its extensive coverage of human movement, recording the trajectories of approximately 180 users in Beijing, China, over more than four years. Despite its detailed representation of diverse activities, the dataset's relatively small user base limits its ability to generalize urban mobility patterns. Complementing this, the YJ-Mob 100k dataset \cite{yabe2024yjmob100k} offers a large-scale perspective with mobility data from 100,000 individuals over 75 days, collected via mobile phones on a metropolitan scale. The dataset's anonymized, grid-based structure facilitates the analysis of broader human mobility trends, although it lacks the granularity to infer individual visit patterns. Additionally, other trajectory datasets, such as those capturing taxi movements in Beijing \cite{yuan2010t} and San Francisco \cite{PSG09}, and bus trajectories in Rio de Janeiro \cite{DC18}, provide insights into traffic dynamics but are less effective in understanding human mobility due to the variability of vehicle passengers.

Synthetic trajectory datasets with high fidelity offer a valuable solution to data accessibility challenges, especially when dealing with privacy concerns, proprietary barriers, and inconsistencies in real-world datasets. The Patterns of Life Simulation \cite{kim2020location} generates city-level human mobility data by utilizing OpenStreetMap to model agents moving between various locations such as homes, workplaces, restaurants, and recreational sites. The simulation is driven by Maslow's hierarchy of needs \cite{maslow1943theory}, influencing agents' decisions and interactions based on physiological, safety, and social needs. A significant example of this application is presented in \cite{amiri2023massive}, where a massive dataset of over 1.5 terabytes was generated, comprising more than 22 billion trajectory locations, 423 million check-ins, and 1.7 billion social links. The simulation's extensive parameters allow for detailed modeling of agents' behaviors, though the impact of these parameters across different global regions remains unclear.

\vspace{-0.2cm}
\section{Data Generation Methodology}\vspace{-0.1cm}
\label{sec:methodology}
This section outlines the data generation process, including the description of anomaly scenarios, simulation of these scenarios, and the steps for data processing and labeling.

\vspace{-0.3cm}
\subsection{Anomaly Scenarios}\vspace{-0.1cm}
\label{sec:anomaly-scenarios}
Anomaly behavior refers to an observation that deviates significantly from the majority of other observations. Such behavior can manifest differently across various environments and applications~\cite{hawkins1980identification}. In this study, we identified four distinct types of anomalies in the simulations: "hunger", "interest", "social", and "work", each of which was further categorized based on intensity levels as "red," "orange," or "yellow." Additionally, new anomalies could be injected in various ways, including random assignment, transmission from other agents, or infection from visited locations.\vspace{-0.15cm}

\subsubsection{Different Types of Anomalies}

In this study, we inject four distinct types of anomalies into the dataset to simulate varied abnormal behavior patterns in agents. These anomalies are classified based on their deviation from standard behavioral patterns, with the possibility of overlapping characteristics among the categories:

\vspace{-0.21cm}
\paragraph{\textbf{Hunger Anomalies}} exhibit an abnormally heightened appetite, leading them to seek food more frequently than usual. In the Patterns of Life Simulation~\cite{zufle2023urban} that our dataset generation is based on, agents have a Food Need that requires them to visit a place where food is available once their Food Need becomes critical. An agent-specific attribute Appetite controls the level at which their food needs increase over time. For hunger anomalies, this Appetite attribute is increased.
This increased drive compels them to visit restaurants or return home to eat at a significantly higher rate than their typical behavior. As a result, their heightened need for sustenance disrupts their regular routines, causing notable deviations in their daily patterns, especially related to their work schedule by requiring agents to leave for to eat and return to resume work.

\vspace{-0.21cm}
\paragraph{\textbf{Interest Anomalies}} undergo significant shifts in their group affiliations and recreational preferences when they become anomalies. These changes drive them to modify their social interactions and choose different recreational sites than they would under normal circumstances. As these anomalies integrate into new social circles, they can create a ripple effect within their broader social network, potentially influencing the behaviors and preferences of other connected individuals.

\vspace{-0.21cm}
\paragraph{\textbf{Social Anomalies}} exhibit a notable deviation in their decision-making process regarding the selection of recreational sites. Agents in the Patterns of Life Simulation~\cite{zufle2023urban} have a Love Need that causes them to visit recreational sites to meet old and new friends and evolve their social network.
Unlike their usual behavior, which might prioritize proximity or familiarity, these agents begin to choose random and often unfamiliar locations for social activities. This randomness introduces unpredictability into their social behavior, making it difficult to anticipate their movements and interactions within the simulated environment.

\vspace{-0.21cm}
\paragraph{\textbf{Work Anomalies}} experience a substantial disruption in their professional routines, ceasing to attend work depending on the intensity of their anomaly status during the relevant period. This absence not only impacts the anomaly's activities but may also have broader implications, potentially affecting the behavior of other agents who depend on the anomaly's presence for their own workplace-related activities.


\subsubsection{Intensity Levels}

To vary the degree of behavior change of anomalous agents and the corresponding level of difficulty in detecting them, we use three levels of behavior change for each anomaly scenario. These deviations are coded by three colors: red, orange, and yellow, signifying varying degrees of behavioral anomaly:
\vspace{-0.21cm}
\paragraph{\textbf{Red Anomalies}} represent the most extreme category of anomaly behavior, characterized by substantial and radical departures from their established patterns. %
Red hunger anomalies will have an extremely high appetite, requiring them to frequently visit a place where they can eat (their home, a restaurant, or a recreational site). Red work anomalies will stop working entirely on 100\% of days and red social anomalies will always choose a random recreational site to visit instead of their (normal) favorite sites.

\vspace{-0.21cm}
\paragraph{\textbf{Orange Anomalies}} demonstrate moderate deviations from expected behaviors. Although these deviations are less severe than those observed in Red Anomalies, Orange Anomalies can still markedly influence their day-to-day activities and the dynamics of their interactions and environments.
Orange Hunger Anomalies will have an increased appetite albeit not as extreme as Red Hunger Anomalies. Orange Work Anomalies will skip going to work 50\% of times (decided independently each time an agent would normally go to work), and Orange Social Anomalies will choose a random recreational site in 50\% of times (decided independently each time an agent decides to visit a recreational site) and use their normal recreational site choice otherwise.

\vspace{-0.21cm}
\paragraph{\textbf{Yellow Anomalies}} exhibit the least severe form of deviation, with only minor alterations from typical behavior patterns. While these changes are subtle and may not cause immediate disruptions to their routines or have a pronounced impact on their environment, they can gradually affect interactions and subtly alter the ambient dynamics over time.

\subsubsection{Mechanisms of Anomaly Injection}\label{subsec:injections}

In the simulation environment, anomalies can be injected through three distinct mechanisms, each designed to model different aspects of anomaly behavior. These mechanisms provide researchers with a flexible framework to simulate anomaly scenarios and evaluate the effectiveness of anomaly detection algorithms in identifying abnormal behavior patterns. The three mechanisms are as follows:
\vspace{-0.21cm}
\paragraph{\textbf{Central Injection}} involves the random selection of agents to be designated as anomalies through a centralized manipulation process. This approach permits researchers to systematically inject anomalies into the simulation environment, facilitating a controlled examination of anomaly behavior patterns and their effects on the interactions and dynamics of the agents within the environment.
\vspace{-0.21cm}
\paragraph{\textbf{Infectious Disease Spread}} simulates the spread of anomalous behaviors among agents using a Susceptible-Exposed-Infectious-Recovered disease model (SEIR)~\cite{he2020seir}. Starting with a few initially infected individuals selected at random using the centralized injection mechanism. All other agents start susceptible. Infected agents may infect susceptible agents through co-location, that is, by being at the same point of interest. Newly infected agents become exposed for a duration chosen uniformly between zero and seven days. Exposed agents can not (yet) infect other agents. After the exposed period, agents immediately become infectious for a duration chosen uniformly in 7-14 days. After this, agents become recovered and cannot be infected again. 
\vspace{-0.21cm}
\paragraph{\textbf{Location-Based Infection}} introduces anomalies through agents' exposure to a specific location. Agents visiting these places have a chance of adopting abnormal behavior patterns. Similar to the infectious disease model, agents become latent (exposed) for 0-7 days. This duration is added to avoid trivial detection of the location causing anomalies. Afterwards, agents adopt anomalous behavior for 7-14 days. In this model, agents do not infect each other and can only be infected through exposure to the infected location similar to the 1854 London Cholera outbreak~\cite{snow1855mode}.



\vspace{-0.3cm}
\subsection{Simulation of Anomalies}
\label{sec:simulation}

To simulate the data, we employ a Patterns of Life Simulation and inject a set of agents specifically designed to exhibit anomaly behavior. The source code of the simulation is available at \GitHubRepo{}

\subsubsection{Pattern of the simulation}


To synthetically generate data, we employ a Patterns of Life Simulation as described in~\cite{kim2020location,zufle2023urban} to simulate individual movement within urban environments. This simulation utilizes OpenStreetMap to create realistic city-level human mobility data by modeling agents navigating and interacting with various locations, including homes, workplaces, restaurants, and recreational sites. The simulation is uniquely informed by Maslow’s hierarchy of needs~\cite{maslow1943theory}, which influences the agents’ decisions and movements based on a range of physiological, safety, and social factors. By integrating these needs into the agents' decision-making processes, the simulation achieves a higher level of behavioral realism, offering a nuanced understanding of how different external and internal stimuli affect human mobility patterns in urban settings.

To simulate the data, we input a comprehensive set of parameters that define the characteristics of both the agents and the environment. These parameters encompass the number of agents, the city map, the quantity and distribution of recreational sites, among other critical factors. Based on these inputs, the simulation generates a dynamic set of agents that navigate the city, with their movements being influenced by a combination of factors, including their individual needs, interactions with other agents, affiliations with specific interest groups, and environmental conditions. The output of the simulation consists of detailed logs capturing the agents' locations, timestamps, types of activities, and other pertinent information, which are essential for further analysis and interpretation of the simulated behaviors and interactions within the urban environment.

\subsubsection{Simulating Anomaly Scenarios}
To simulate anomaly scenarios, we inject a set of agents specifically designed to exhibit anomaly behavior, as detailed in Section~\ref{sec:anomaly-scenarios}. These agents are programmed to intentionally deviate from typical patterns of life by engaging in activities that are uncommon or atypical for the general population. Anomalies are generated through three distinct mechanisms: central manipulation, the propagation of anomalous behavior, and location-based deviations. These mechanisms are strategically employed to model various facets of anomaly behavior and to assess the effectiveness of detection algorithms in identifying abnormal patterns.

Each mechanism is further classified into three intensity levels—coded as red, orange, and yellow—which represent the severity of deviation from normative behavior across four categories of anomalies: hunger, interest, social, and work. The red level signifies the most severe deviations, while orange and yellow indicate moderate and mild deviations, respectively. In the following sections, we provide an in-depth description of each type of anomaly and their associated intensity levels, while emphasizing that the methods of injection align with the mechanisms outlined in Section~\ref{sec:anomaly-scenarios}. This approach ensures a comprehensive evaluation of how well the detection algorithms can identify and differentiate between various types and severities of anomaly behavior. In the following, we describe the different types of anomalies and their associated intensity levels in detail.
\vspace{-0.212cm}
\paragraph{\textbf{Hunger Anomalies}} increase their food need, leading to more frequent visits to restaurants or home kitchens. This is achieved by adjusting two key parameters: (1) the time after eating when they begin to feel hungry and (2) the rate at which their hunger increases.
At the red anomaly level, the time it takes for them to start feeling hungry is reduced to zero, meaning they are perpetually hungry, and the rate at which their hunger increases is tripled, causing a rapid decline in fullness and necessitating constant food consumption.
At the orange anomaly level, the time before they start to feel hungry is cut in half, and their rate of hunger increase is doubled. This results in a significant increase in the frequency of their food consumption, though not as extreme as the red level.
At the yellow anomaly level, the time before they start feeling hungry is shortened to 75\% of their normal value, and the rate at which they become hungry is increased by 1.5 times. This causes a moderate increase in their food consumption, leading to more frequent meals but still allowing for some periods of satiety.
\vspace{-0.212cm}
\paragraph{\textbf{Interest Anomalies}} exhibit dynamic behavior in their social interactions and recreational site visitation patterns by periodically shifting their focus from one interest group to another. This behavior is classified into different levels based on the frequency of interest changes, which correspondingly influences the degree of variability in their activities. At the red anomaly level, individuals change their interest daily, resulting in a high degree of variability in both their social interactions and site visits. The orange anomaly level is characterized by an interest change every other day, leading to moderate variability. Lastly, the yellow anomaly level involves a weekly change in interest, causing a mild degree of variability in their interactions and activities. This classification allows for a nuanced understanding of how different levels of interest variability impact the agents' behavior and the challenges associated with detecting such anomalies.
\vspace{-0.212cm}
\paragraph{\textbf{Social Anomalies}} demonstrate unique patterns in their selection of recreational sites by occasionally opting for different locations, even when a more appealing site is available nearby. This behavior can be categorized into various levels based on the likelihood of site changes, which directly impacts the degree of variability in their visitation patterns. At the red anomaly level, agents randomly select a recreational site 100\% of the time, leading to a very high degree of variability in their site visits. The orange anomaly level is characterized by random site selection 50\% of the time, resulting in a moderate level of variability. Finally, at the yellow anomaly level, random site selection occurs 20\% of the time, leading to only mild variability in their site visits. These different levels of anomaly behavior reflect the agents' varying tendencies to deviate from more predictable visitation patterns.
\vspace{-0.212cm}
\paragraph{\textbf{Work Anomalies}} display atypical work patterns by occasionally ceasing work during planned work periods. This behavior is categorized into different levels based on the frequency of these interruptions, which directly affects the degree of variability in their work schedules. At the red anomaly level, individuals cease working 100\% of the time when they plan to work, resulting in an extremely high degree of variability in their work patterns. The orange anomaly level is defined by work interruptions occurring 50\% of the time, leading to a moderate degree of variability. Finally, at the yellow anomaly level, work interruptions occur 20\% of the time, causing only mild variability in their work schedules. This classification provides a nuanced understanding of how varying levels of work interruptions impact agents' behavior and the challenges associated with identifying such anomalies.


\subsubsection{Simulation Specifications}
After implementing and running the simulation with the specified parameters, we collected data on agents' trajectories, staypoints, and social connections. The trajectory data includes the locations and timestamps of agents, which are essential for analyzing movement patterns and identifying potential anomalies. The staypoint data captures stationary locations, including the type of location and social links between agents, providing valuable insights into their interactions.

In this study, we simulated 1,000 agents in the cities of Atlanta and Berlin. The first four weeks served as a warm-up period, followed by four weeks representing normal life conditions. The anomaly period began after eight weeks of simulation time.
During the anomaly period, we injected central manipulation, infection-based, and location-based anomalies in different simulation setups. Central manipulation was carried out over four weeks, involving 120 agents, while infection and location-based anomalies were simulated over 12 weeks. Initially, 10 agents were infected, with the infection rate varying across scenarios ([0.01, 0.05, 0.1, 0.5, 1]), determining the number of additional agents affected. Infected agents recovered after one week and before two weeks and were not reinfected later. For location-based anomalies, the infection spread to either the nearest location to a randomly assigned agent or the most popular recreational site in the city. The infection rate for these locations matched that used in the infectious disease simulation, ensuring consistency in modeling disease spread dynamics.

\vspace{-0.3cm}
\subsection{Data Processing and Labeling}
\label{sec:data-processing}
We provide three distinct datasets for each simulation scenario: trajectory data, stay-point data, and social links. The trajectory dataset captures the location history of each agent at 5-minute intervals, offering detailed insights into their movement patterns. The stay-point dataset includes information on the locations where agents linger, detailing the time of arrival, venue types, and corresponding coordinates. The social links dataset documents the social connections between agents, including agent IDs, friend IDs, and the time of these relationships. These datasets are meticulously generated using our proposed methodology and are well-suited for a variety of applications, including anomaly detection, pattern-of-life analysis, and social network analysis.

To ensure compatibility and standardization, we have converted the coordinate system of the logs to the WGS84 standard. This conversion guarantees global coverage and high accuracy of location data, which is essential for applications such as mapping, navigation, and geospatial analysis. The WGS84 standard is universally adopted in GPS devices and mapping applications, making it the preferred choice for standardizing location data. By standardizing our datasets to WGS84, we ensure seamless integration with other geospatial datasets and applications, facilitating interoperability and analysis across diverse platforms and systems. This compatibility significantly enhances the utility of our datasets in various geospatial and analytical applications.

Following preprocessing, we labeled the data to identify and classify anomalies based on their type and intensity. During the labeling process, normal life activities were assigned a label of '0', while anomalies were labeled according to their type and intensity. Anomaly types were encoded as integers representing specific categories: hunger (1), work (2), social (3), and interest (4). The intensity of each anomaly was similarly encoded with values ranging from 1 to 3, representing increasing levels of severity: 1 for severe (red), 2 for moderate (orange), and 3 for mild (yellow). The final label for each anomaly was constructed by concatenating the anomaly type and intensity values. For example, a hunger-related anomaly with an intensity of 2 would be labeled as '12'. This nuanced labeling approach not only identifies the presence of an anomaly but also distinguishes its type and severity, thereby enhancing the model's ability to detect and classify anomalies with greater accuracy across various contexts.

\vspace{-0.2cm}
\section{Dataset Description}
\label{sec:description}
In this section, we provide a detailed description of the generated datasets available at \url{https://osf.io/dg6t3/}. We first present the specifications of the datasets in terms of the structure and the statistics including the number of agents, the number of GPS locations, the number of stay points, and the number of social connections, among others. We also provide an in-depth analysis of the generated datasets, utilizing both spatial and non-spatial data analysis techniques to visualize the anomaly patterns.

\vspace{-0.3cm}
\subsection{Datasets Specifications}
\label{sec:description-datasets}
\subsubsection{Structure of the Datasets}

The datasets are systematically organized into three main directories, each corresponding to a distinct method of anomaly injection: (1) centralized manipulation mechanisms, (2) infectious disease spread, and (3) location-based spread anomalies. Within each main directory, sub-directories represent various simulation scenarios, differentiated by geographical region, simulation parameters, and the specific type of anomaly simulated. The centralized manipulation mechanism involves random injections of anomalies to agents, while the infectious disease spread scenario models the transmission of disease through a population using a standard infectious disease framework. In the location-based spread scenario, anomalies are triggered when agents interact with specific locations. The simulated data cover two geographical regions—Atlanta, GA, and Berlin, Germany. Furthermore, the sub-directories within each directory categorize the datasets by additional parameters, including the type of anomaly being simulated, such as hunger, social, work, interests, or a combination of these factors.

Each sub-directory contains a set of files that provide the actual dataset, ground truth labels for the anomalies, and meta information about the data generation/processing. The data files are stored in CSV format and compressed using the ZIP format, while the ground truth labels and the meta information are stored in a JSON file. The data files include trajectories, stay-points, and social links, which offer a detailed view of agent movements, locations visited, and social interactions. Each dataset is divided into training and testing phases, with the training data typically spanning a four-week period and the testing data varying based on the type of anomalies being studied. The "labels.json" file in each sub-directory provides the ground truth for the anomalies, detailing the type and severity of each agent's anomaly behavior including agent ID, label, and the start and end times of the anomaly period. The "info.json" file contains additional information about the dataset, such as the number of agents, start and end times of the train and test data, the number of GPS locations, stay-points, and social connections, specific simulation parameters, among others.

\vspace{-0.23cm}
\subsubsection{Dataset Statistics}

Statistics of the generated datasets such as overall specification, number of data points, size of the generated datasets, agent to agent and location to agent infectious disease anomalies are presented in \reftab{tab:number-of-data-points}, \reftab{tab:size-of-datasets}, \reftab{tab:statistics-infectious-disease} and \reftab{tab:statistics-location-based}, respectively. The datasets were generated to model various scenarios using different injection methods, including Central Manipulation Assignment (CMA), Infectious Disease Spread (IDS), and Location-Based Spread (LBS). These methods were employed to simulate realistic movement patterns and social interactions within a population, capturing essential data types such as Trajectories, Staypoints, and Social Links. The resulting datasets were subsequently divided into training and testing sets to facilitate model training and evaluation under diverse conditions.

\reftab{tab:overview} provides an overall specification of the simulated data for three different injection methods: Central Manipulation Assignment (CMA), Infectious Disease Spread (IDS), and Location-Based Spread (LBS). Each method involves three distinct periods: Pre-train, Train, and Test. The Pre-train and Train periods are consistent across all methods, each lasting 4 weeks. However, the Test period differs, with CMA having a 4-week duration, while IDS and LBS extend over 12 weeks. Notably, the Pre-train period is used to warm up the agents, allowing them to settle into their simulated lives, such as establishing jobs and social networks.
The number of anomalies detected varies among the methods; CMA has a fixed 120 anomalies, IDS ranges widely from 10 to 966 anomalies, and LBS ranges from 3 to 463 anomalies. 

\begin{table}[t]
    \centering
    \caption{Overall Specification of the Simulated data;
    CMA: Central Manipulation Assignment, IDS: Infectious Disease Spread, LBS: Location-Based Spread;}
    \vspace{-0.3cm} 
    \label{tab:overview}
    \begin{tabular}{|l|l|l|l|l|}
        \hline
        & Pre-train  & Train  & Test  & \#Anomalies \\ \hline
        CMA & 4 weeks & 4 weeks & 4 weeks & 120 \\ \hline
        IDS & 4 weeks & 4 weeks & 12 weeks & (10-966) \\ \hline
        LBS & 4 weeks & 4 weeks & 12 weeks & 3-463 \\ \hline
    \end{tabular}
    \vspace{-0.3cm}
\end{table}

\begin{table}[t]
    \centering
    \caption{Number of Data Entry Points in the Generated Datasets for the Different Infection Methods; CMA: Central Manipulation Assignment, IDS: Infectious Disease Spread, LBS: Location-Based Spread;
        M: Million, K: Thousand; (The numbers may vary slightly for different scenarios)}
        \vspace{-0.3cm}
    \label{tab:number-of-data-points}
    \begin{tabular}{|l|c|c|c|c|c|c|}
        \hline
                & \multicolumn{2}{c|}{Trajectory} & \multicolumn{2}{c}{Staypoint} & \multicolumn{2}{|c|}{Social Links}                       \\ \hline

        Dataset & Train                           & Test                          & Train                              & Test & Train & Test \\ \hline
        CMA     & 8M                              & 8M                            & 158K                               & 150K & 562K  & 745K \\ \hline
        IDS     & 8M                              & 24M                           & 158K                               & 450K & 562K  & 2M   \\ \hline
        LBS     & 8M                              & 24M                           & 158K                               & 440K & 562K  & 2M   \\ \hline
    \end{tabular}
    \vspace{-0.3cm}
\end{table}

\reftab{tab:number-of-data-points} provides an overview of the number of data entry points in the generated datasets for different infection methods: Central Manipulation Assignment (CMA), Infectious Disease Spread (IDS), and Location-Based Spread (LBS). The datasets are categorized into three types: Trajectory, Staypoint, and Social Links, each with distinct entries for training and testing phases. For Trajectory data, both CMA and IDS/LBS methods have 8 million data points for training, while the test sets differ, with IDS and LBS having 24 million data points compared to 8 million for CMA. The Staypoint data shows a similar pattern in the training sets (158 K), with increasing data points in the test sets from CMA (150 K) to IDS (450 K) and LBS (440 K). Social Links data also follows a similar trend, with CMA having 562 K training data points and 745 K test data points, whereas IDS and LBS have the same number of training points but significantly more test data points at 2 million. This variation in data points highlights the differing scales and complexities of the datasets generated by each injection method, reflecting their potential use in simulating different scenarios.

\begin{table}[!ht]
    \centering
    \caption{Size of the Generated Datasets for the Different Injection Methods; CMA: Central Manipulation Assignment, IDS: Infectious Disease Spread, LBS: Location-Based Spread;
        MB: Megabyte, GB: Gigabyte; (The sizes may vary slightly for different scenarios)}
        \vspace{-0.3cm}
    \label{tab:size-of-datasets}
    \begin{tabular}{|l|c|c|c|c|c|c|}
        \hline
                & \multicolumn{2}{c|}{Trajectory} & \multicolumn{2}{c}{Staypoint} & \multicolumn{2}{|c|}{Social Links}                       \\ \hline

        Dataset & Train                           & Test                          & Train                              & Test & Train & Test \\ \hline
        CMA     & 514MB                           & 515MB                         & 11MB                               & 11MB & 15MB  & 21MB \\ \hline
        IDS     & 514MB                           & 1GB                           & 11MB                               & 33MB & 15MB  & 64MB \\ \hline
        LBS     & 514MB                           & 1GB                           & 11MB                               & 32MB & 15MB  & 63MB \\ \hline
    \end{tabular}
    \vspace{-0.4cm}
\end{table}

\reftab{tab:size-of-datasets}  summarizes the size of generated datasets for different infection methods: Central Manipulation Assignment (CMA), Infectious Disease Spread (IDS), and Location-Based Spread (LBS). The datasets are divided into three categories—Trajectory, Staypoint, and Social Links—with sizes provided for both training and testing datasets. For the Trajectory datasets, the training size is consistent across all methods at 514 MB, while the test size for IDS and LBS is significantly larger (1 GB) compared to CMA (515 MB). In the Staypoint datasets, the training size remains constant at 11 MB, but the test sizes are 11 MB for CMA to 33 MB for IDS and 32 MB for LBS. Similarly, for the Social Links datasets, the training size is 15 MB for all methods, but the test sizes vary, with CMA at 21 MB, IDS at 64 MB, and LBS at 63 MB. These variations indicate differences in data density and complexity among the injection methods, reflecting the nature of the scenarios they simulate.

\begin{table}[t]
    \centering
    \caption{Statistics of the Generated Datasets for Infectious Disease Spread Injection Method in Atlanta, GA; \#E: Number of Exposed, \#I: Number of Infected, \#R: Number of Recovered} \vspace{-0.3cm}
    \label{tab:statistics-infectious-disease}
    \begin{tabular}{|c|c|c|c|c|c|}
        \hline
        Infectious Rate & Anomaly Type & \#E & \#I & \#R \\ \hline
        0.01            & combined     & 7   & 45  & 45  \\ \hline
        0.01            & hunger       & 1   & 11  & 11  \\ \hline
        0.01            & interest     & 1   & 11  & 11  \\ \hline
        0.01            & social       & 0   & 10  & 10  \\ \hline
        0.01            & work         & 1   & 11  & 11  \\ \hline
        0.05            & combined     & 39  & 77  & 77  \\ \hline
        0.05            & hunger       & 22  & 32  & 32  \\ \hline
        0.05            & interest     & 24  & 34  & 34  \\ \hline
        0.05            & social       & 27  & 37  & 37  \\ \hline
        0.05            & work         & 24  & 34  & 34  \\ \hline
        0.10            & combined     & 699 & 737 & 736 \\ \hline
        0.10            & hunger       & 438 & 420 & 331 \\ \hline
        0.10            & interest     & 686 & 692 & 682 \\ \hline
        0.10            & social       & 666 & 672 & 651 \\ \hline
        0.10            & work         & 735 & 744 & 732 \\ \hline
        0.50            & combined     & 937 & 975 & 975 \\ \hline
        0.50            & hunger       & 968 & 978 & 978 \\ \hline
        0.50            & interest     & 953 & 963 & 963 \\ \hline
        0.50            & social       & 963 & 973 & 973 \\ \hline
        0.50            & work         & 962 & 972 & 972 \\ \hline
        1.00            & combined     & 939 & 977 & 977 \\ \hline
        1.00            & hunger       & 971 & 981 & 981 \\ \hline
        1.00            & interest     & 972 & 982 & 982 \\ \hline
        1.00            & social       & 973 & 983 & 983 \\ \hline
        1.00            & work         & 966 & 976 & 976 \\ \hline
        
    \end{tabular}
    \vspace{-0.5cm}
\end{table}

\paragraph{Infectious Disease Spread}
The datasets generated using the Infectious Disease Spread method in Atlanta, GA, are meticulously detailed in \reftab{tab:statistics-infectious-disease} This table presents key statistics, including the number of exposed (E), infected (I), and recovered (R) individuals under varying conditions. The data is organized according to different infectious rates and categorized by anomaly types, such as combined, hunger, interest, social, and work, to capture the impact of these factors on the spread of the disease.

\paragraph{Location-Based Spread}
The datasets generated using the Location-Based Injection method in Atlanta, GA, are detailed in~\reftab{tab:statistics-location-based}. This table presents the number of exposed (E), infected (I), and recovered (R) individuals, providing insights into how the spread of infection varies based on different infection probabilities, selection methods of the source recreational site (random or popular), and anomaly types (combined, hunger, interest, social, work).
The inclusion of selection methods offers an additional layer of analysis, allowing for a comparison between scenarios where locations are chosen either randomly or based on popularity. The data shows that higher infectious rates generally lead to an increased number of exposed, infected, and recovered individuals, as expected. Moreover, the choice of selection method and anomaly type significantly influences the spread of infection. For instance, at higher infectious rates, popular locations tend to result in higher numbers of infections, reflecting the potential for rapid disease spread in densely populated or frequently visited areas.

\begin{table}[b]
    \centering
    \caption{Statistics of the Generated Datasets for Location-Based Injection Method in Atlanta, GA; \#E: Number of Exposed, \#I: Number of Infected, \#R: Number of Recovered}
    \vspace{-0.3cm}
    \label{tab:statistics-location-based}
    \small\begin{tabular}{|c|c|c|c|c|c|}
        \hline
        Rate & Selection & Anomaly Type & \#E & \#I & \#R \\ \hline
        0.01 & random    & combined     & 14  & 13  & 11  \\ \hline
        0.01 & random    & hunger       & 7   & 6   & 6   \\ \hline
        0.01 & random    & interest     & 3   & 3   & 3   \\ \hline
        0.01 & random    & social       & 4   & 3   & 3   \\ \hline
        0.01 & random    & work         & 7   & 6   & 6   \\ \hline
        0.01 & popular   & combined     & 6   & 6   & 6   \\ \hline
        0.01 & popular   & hunger       & 6   & 6   & 6   \\ \hline
        0.01 & popular   & interest     & 6   & 6   & 5   \\ \hline
        0.01 & popular   & social       & 6   & 6   & 5   \\ \hline
        0.01 & popular   & work         & 6   & 6   & 5   \\ \hline
        0.05 & random    & combined     & 45  & 41  & 36  \\ \hline
        0.05 & random    & hunger       & 27  & 26  & 26  \\ \hline
        0.05 & random    & interest     & 20  & 18  & 17  \\ \hline
        0.05 & random    & social       & 23  & 21  & 20  \\ \hline
        0.05 & random    & work         & 24  & 24  & 24  \\ \hline
        0.05 & popular   & combined     & 23  & 22  & 21  \\ \hline
        0.05 & popular   & hunger       & 23  & 22  & 21  \\ \hline
        0.05 & popular   & interest     & 26  & 26  & 25  \\ \hline
        0.05 & popular   & social       & 25  & 25  & 23  \\ \hline
        0.05 & popular   & work         & 22  & 22  & 21  \\ \hline
        0.10 & random    & combined     & 86  & 81  & 79  \\ \hline
        0.10 & random    & hunger       & 48  & 47  & 44  \\ \hline
        0.10 & random    & interest     & 24  & 22  & 16  \\ \hline
        0.10 & random    & social       & 34  & 29  & 27  \\ \hline
        0.10 & random    & work         & 51  & 50  & 46  \\ \hline
        0.10 & popular   & combined     & 45  & 44  & 42  \\ \hline
        0.10 & popular   & hunger       & 45  & 44  & 42  \\ \hline
        0.10 & popular   & interest     & 46  & 45  & 40  \\ \hline
        0.10 & popular   & social       & 40  & 39  & 36  \\ \hline
        0.10 & popular   & work         & 47  & 45  & 42  \\ \hline
        0.50 & random    & combined     & 324 & 309 & 285 \\ \hline
        0.50 & random    & hunger       & 198 & 195 & 179 \\ \hline
        0.50 & random    & interest     & 83  & 81  & 68  \\ \hline
        0.50 & random    & social       & 94  & 91  & 81  \\ \hline
        0.50 & random    & work         & 215 & 208 & 187 \\ \hline
        0.50 & popular   & combined     & 183 & 176 & 157 \\ \hline
        0.50 & popular   & hunger       & 183 & 176 & 157 \\ \hline
        0.50 & popular   & interest     & 174 & 170 & 154 \\ \hline
        0.50 & popular   & social       & 168 & 164 & 154 \\ \hline
        0.50 & popular   & work         & 172 & 168 & 154 \\ \hline
        1.00 & random    & combined     & 472 & 463 & 437 \\ \hline
        1.00 & random    & hunger       & 287 & 279 & 257 \\ \hline
        1.00 & random    & interest     & 123 & 118 & 108 \\ \hline
        1.00 & random    & social       & 157 & 149 & 134 \\ \hline
        1.00 & random    & work         & 303 & 293 & 267 \\ \hline
        1.00 & popular   & combined     & 269 & 262 & 237 \\ \hline
        1.00 & popular   & hunger       & 269 & 262 & 237 \\ \hline
        1.00 & popular   & interest     & 275 & 268 & 247 \\ \hline
        1.00 & popular   & social       & 272 & 263 & 238 \\ \hline
        1.00 & popular   & work         & 264 & 257 & 236 \\ \hline
    \end{tabular}
    \vspace{-0.4cm}
\end{table}

\begin{figure*}[t]
\centering
\begin{subfigure}{0.32\linewidth}
\centering
\includegraphics[width=\linewidth]{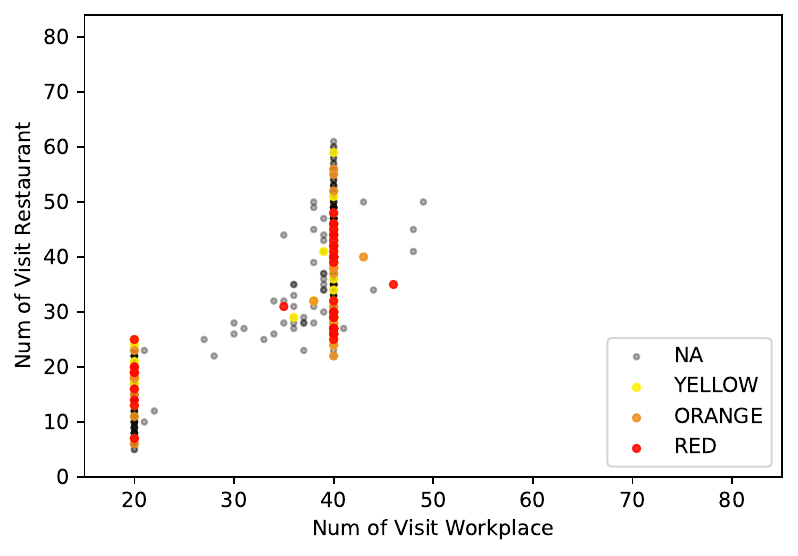}
\vspace{-0.6cm}
\caption{Hunger Anomalies: Train}
\label{fig:wr_train_hunger}
\end{subfigure} \rulesep
\begin{subfigure}{0.32\linewidth}
\centering
\includegraphics[width=\linewidth]{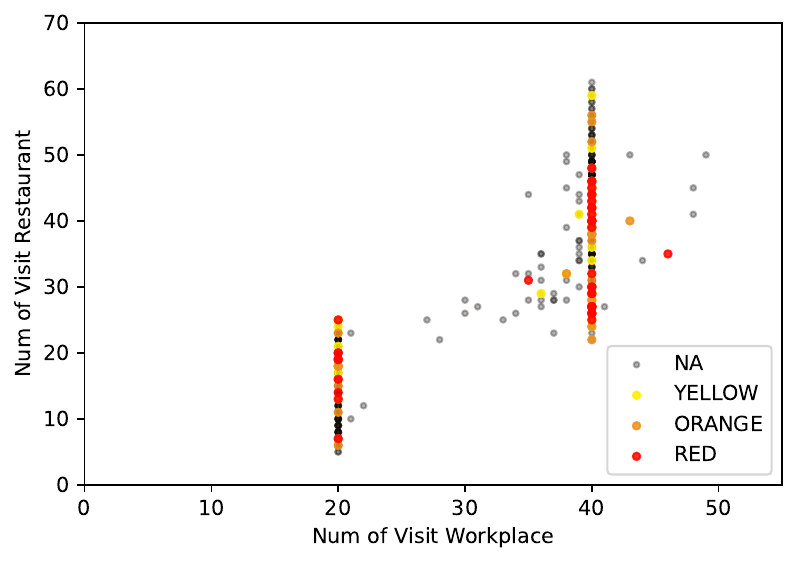}
\vspace{-0.6cm}
\caption{Work Anomalies: Train}
\label{fig:wr_train_work}
\end{subfigure} \rulesep
\begin{subfigure}{0.32\linewidth}
\centering
\includegraphics[width=\linewidth]{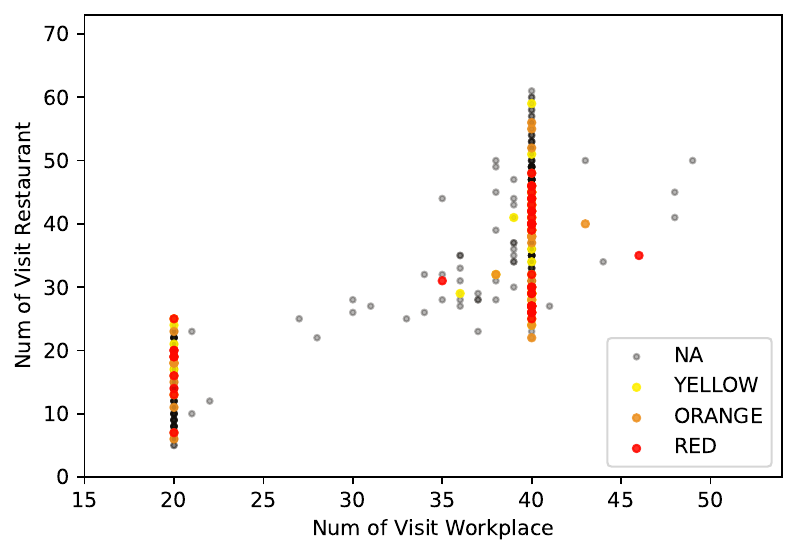}
\vspace{-0.6cm}
\caption{Social Anomalies: Train}
\label{fig:wr_train_social}
\end{subfigure} 
\begin{subfigure}{0.32\linewidth}
\centering
\includegraphics[width=\linewidth]{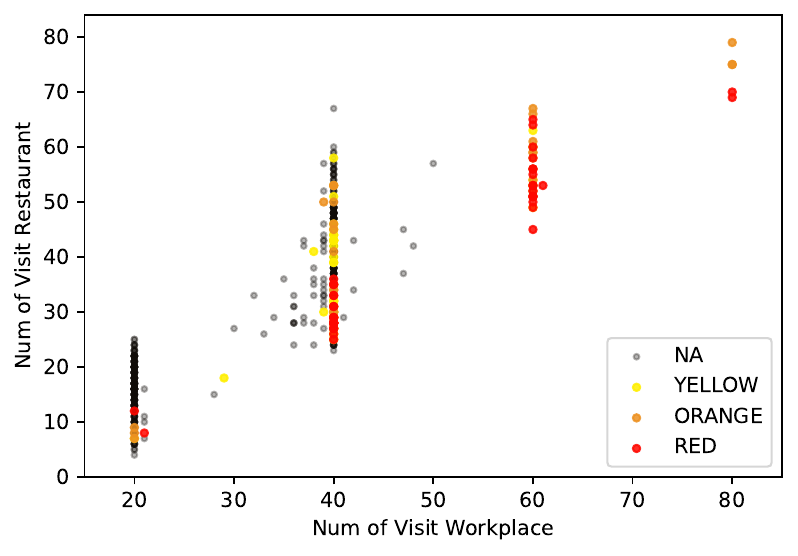}
\vspace{-0.6cm}
\caption{Hunger Anomalies: Test}
\label{fig:wr_test_hunger}
\end{subfigure} \rulesep
\begin{subfigure}{0.32\linewidth}
\centering
\includegraphics[width=\linewidth]{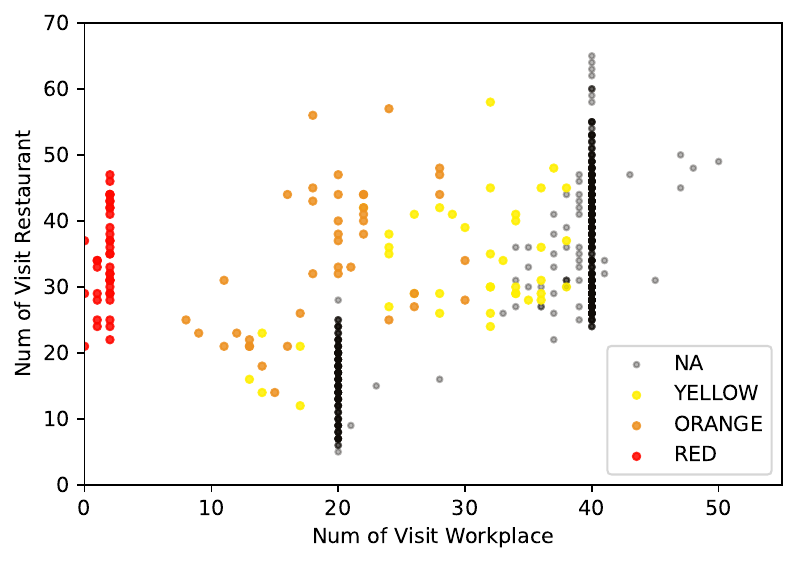}
\vspace{-0.6cm}
\caption{Work Anomalies: Test}
\label{fig:wr_test_work}
\end{subfigure} \rulesep
\begin{subfigure}{0.32\linewidth}
\centering
\includegraphics[width=\linewidth]{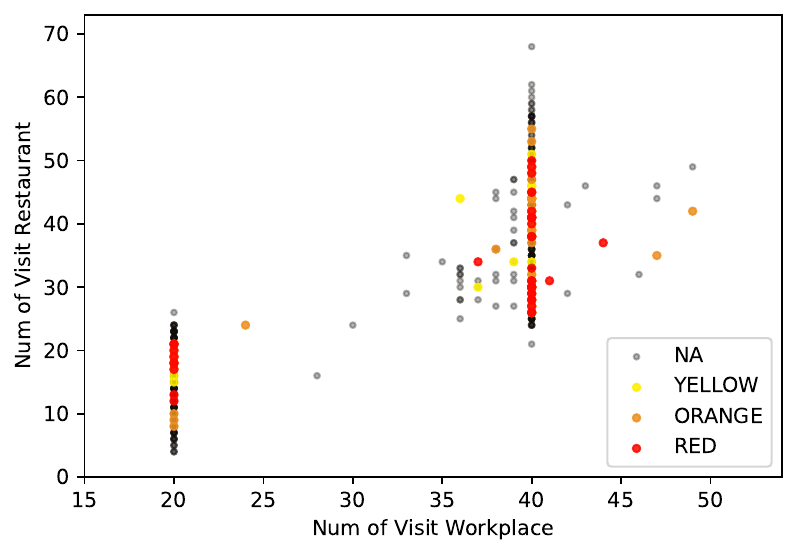}
\vspace{-0.6cm}
\caption{Social Anomalies: Test}
\label{fig:wr_test_social}
\end{subfigure}
\caption{Number of Restaurant Visits versus Number of Workplace Visits for Hunger, Work, and Social Anomalies.}
\label{fig:wr}
\end{figure*}
\vspace{-0.2cm}

\subsection{Datasets Analysis}
\label{sec:description-analysis}
This section provides the basic analysis of the generated datasets for three types of anomalies: hunger, work, and social anomalies to show the existence and capability to distinguish the generated anomaly, as well as the specific patterns in the datasets with the infectious disease model involved. Notably, the interest anomalies need an in-depth social network analysis to be distinguished which is not included in the basic analysis.

\begin{figure*}[t]
\centering
\begin{subfigure}{0.33\linewidth}
\centering
\includegraphics[width=\linewidth]{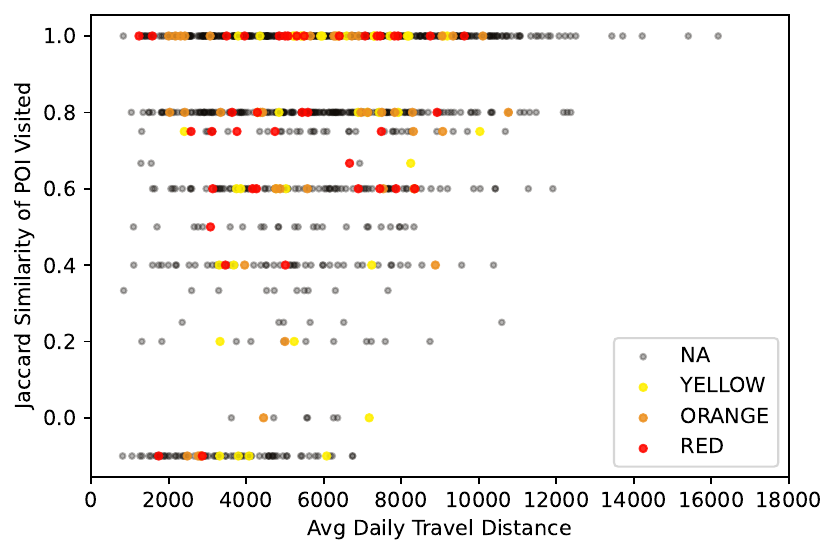}
\caption{Hunger Anomalies}
\label{fig:jd_hunger}
\end{subfigure}
\begin{subfigure}{0.33\linewidth}
\centering
\includegraphics[width=\linewidth]{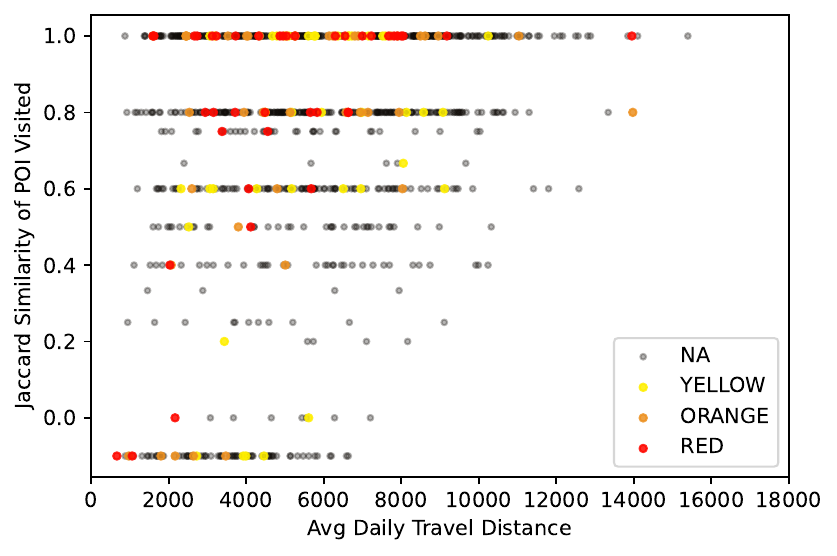}
\caption{Work Anomalies}
\label{fig:jd_work}
\end{subfigure}
\begin{subfigure}{0.33\linewidth}
\centering
\includegraphics[width=\linewidth]{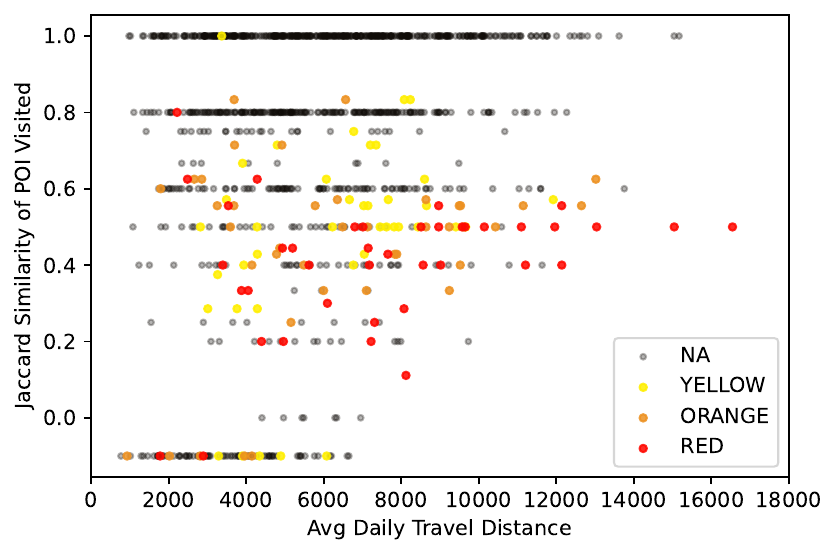}
\caption{Social Anomalies}
\label{fig:jd_social}
\end{subfigure}
\vspace{-0.4cm}
\caption{Jaccard Distance}
\label{fig:jd}
\end{figure*}

\vspace{-0.23cm}
\subsubsection{Anomaly Agents Patterns}
Knowing that the hunger anomalies get hungry faster than normal causing more meals needed per day, and the work anomalies go to work less than normal, for each central anomaly dataset, we plot the number of times an agent visits a restaurant versus a workspace in Figure~\ref{fig:wr}.

The first row of the plot (Figure~\ref{fig:wr_train_hunger} -\ref{fig:wr_train_social}) uses the data from the 28-day training period where all agents behaved normally. During this period, most agents visited a workspace either 20 or 40 times, which is once or twice per workday as there are 20 workdays. The agents visiting a workspace once per workday are the ones who go to work in the morning and stay there without lunch until returning home. The ones who leave to have lunch will correspondingly visit a workspace twice per workday.
Similar plots are created for the 28-day testing period shown in the second row (Figure~\ref{fig:wr_test_hunger} -\ref{fig:wr_test_social}) where the anomalies lived with the designed anomaly behaviors. Considering the hunger anomaly dataset, the anomaly agents get hungry faster than normal requiring more meals per day. These agents will leave their workspace to have meals and return to work. Thus results in additional visits to the workspace after returning from meals. As shown in Figure~\ref{fig:wr_test_hunger}, many of the anomalous agents clearly stand out from the normal agents and, in the most exaggerated condition, they take two more meals than normal, causing four times to visit the workspace per workday or 80 times in the total during the period. 
%
The work anomalies are also distinguishable in this workplace visit to restaurant visit plot as shown in Figure~\ref{fig:wr_test_work}. Red Work Anomalies, which never go to work, are easily discriminated due to having an anomalously low number of work visits. For Orange and Yellow Work Anomalies, this number is not as low, but, in most cases, deviates from the normally observed 20 or 40 work visits allowing to visually discrimate their behavior from normal behavior. 
The social anomalies, however, show the same pattern in these two dimensions, comparing Figure~\ref{fig:wr_test_social} to Figure~\ref{fig:wr_train_social}. To visualize the difference in these social anomalies, we create another graph as in Figure~\ref{fig:jd}, which plots the average daily travel distance in the test period against the Jaccard similarities between the recreational sites they visited during the training and the testing period. Let $\text{Pub}_{train}(i)$ ($\text{Pub}_{test}(i)$) be set of the recreational sites agent $i$ has visited in the training (test) period, then we define the Jaccard similarity between places visited and train and test for agent $i$ as:
\begin{equation*}
    \text{Jaccard}(i) = \frac{|\text{Pub}_{train}(i)\cap \text{Pub}_{test}(i)|}{|\text{Pub}_{train}(i)\cup \text{Pub}_{test}(i)|}
\end{equation*}
If agent $j$ did not visit any recreational site in both the training and testing period, which would yield a Jaccard Similarity of $\frac{0}{0}$, we define $\text{Jaccard}(j) = -0.1$.

Figures~\ref{fig:jd_hunger}-\ref{fig:jd_work} show that Hunger and Work Anomalies do not show any different behavior in terms of distance traveled and Jaccard Similarity. We observe that both normal agents, hunger anomalies, and work anomalies frequently have a Jaccard similarity of $1.0$ (which indicates that the agent visited the exact same sets of recreational sites in train and test) or $0.8$ (indicating the agents visited five recreational sites in total, but one of them was only visited in either train or test).
However, Figure~\ref{fig:jd_social} shows that for Red \emph{Social} Anomalies, their Jaccard Similarity is never at $1.0$ as agents always visit places in Test that they did not visit in Train and their Jaccard Similarity is usually around $0.5$ or below, indicating a large number of new recreational sites visited.
Similar, but less extreme, effects can be observed for Orange and Yellow Social Anomalies.
An interesting case is those agents having a Jaccard Similarity of $-0.1$,  who did not visit any recreational sites during the test or train. Thus, these agents, despite having a change in their behavior, exhibit the exact same trajectory that they would have had without being selected as an anomaly. It may be argued if such agents should be considered as anomalies as their trajectory is identical to the trajectory they would have taken if they were not anomalous.

The interest anomalies are not included in these plots as they do not show differences in all four dimensions. The possible features to distinguish them are left for future research.




\begin{figure}[t]
\centering
\includegraphics[width=\linewidth]{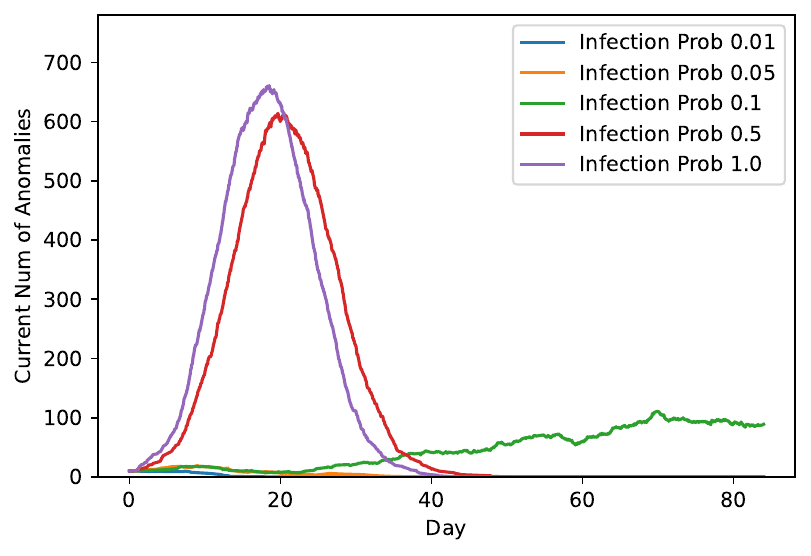}
\caption{Epidemiological Curves for Different Infection Probabilities\vspace{-0.3cm}}
\label{fig:epi_hunger}
\end{figure}

\begin{figure}[t]
\centering
\begin{subfigure}{0.98\linewidth}
\centering
\includegraphics[width=\linewidth]{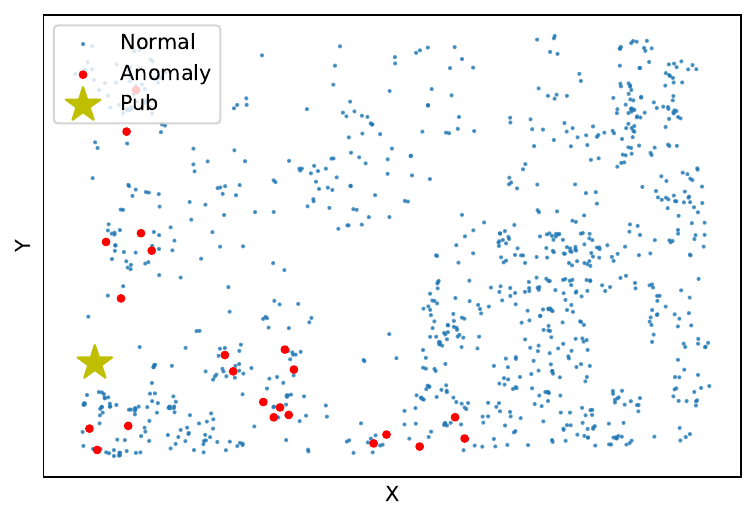}
\vspace{-0.6cm}
\caption{Interest Anomalies}
\label{fig:p2a_interest}
\end{subfigure}
\begin{subfigure}{0.98\linewidth}
\centering
\includegraphics[width=\linewidth]{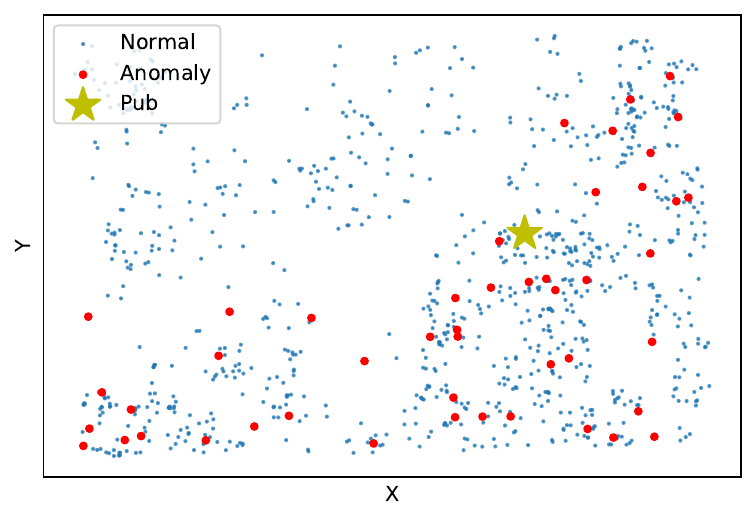}
\vspace{-0.6cm}
\caption{Hunger Anomalies}
\label{fig:p2a_hunger}
\end{subfigure}
\vspace{-0.3cm}
\caption{Spatial Map of Location-Based Anomalies}
\vspace{-0.3cm}
\label{fig:p2a}
\end{figure}

\vspace{-0.23cm}
\subsubsection{Anomaly Injection Patterns}
Besides centrally assigning anomalies at random, we use two other ways to inject anomalies into the running simulation as described in Section~\ref{subsec:injections}: An infectious disease model and a location-based infection scenario. In this section, we evaluate the temporal and spatial patterns of these infection methods.

For the infectious disease model, anomalous agents can infect other agents and make them anomalous. We plot the epidemiological curve for the hunger anomalies as an example in Figure~\ref{fig:epi_hunger}. Each curve in the plot represents a different spreading rate as shown in the legend, and may also represent different detecting difficulties. We note that for the cases having a transmission probability of $0.01$ or $0.05$, for these simulation runs, we did not observe any outbreak beyond the initial infections. For the cases having a transmission probability of $0.5$ or $1.0$, we observe that most of the population of 1000 agents is infected within a few days leading to an extinction of the infectious disease once all agents have recovered and become immune. For the case having a transmission probability of $0.1$, we observe that the infectious disease remains latent among the population without a major outbreak but also without vanishing entirely.

In the case of location-based infectious diseases, the agents can only be infected by visiting an infectious recreational site (such as an infectious well in the John Snow Cholera example). In Figure~\ref{fig:p2a_interest}, the infectious recreational site is located on the edge of the map shown as a yellow star. In this case, all agents who had become anomalies live in the bottom left corner of the map shown as red dots. The phenomenon is due to the design that agents are more likely to choose the nearest pub. If the infectious pub is located in the middle of the map then anomalies may spread in a larger range as in Figure~\ref{fig:p2a_hunger} causing higher difficulties in finding them.


\section{Dataset Regeneration}
\label{sec:regeneration}
To regenerate the data, you need to follow a series of steps, each with its specific instructions and associated files available in the \GitHubRepo{}. The process involves configuring the simulation, running it, and then processing the resulting logs. For each injection method, there is a corresponding configuration script as follows: \texttt{configure\_centralized.py}, \texttt{configure\_infectious.py}, and \texttt{configure\_location.py}. Running these scripts generates a \texttt{.sh} file, which must then be executed. The simulation will run in parallel, so it is important to specify the number of cores to be used in the \texttt{configure*.py} files. Once the simulation is complete, the data is ready for processing, which can be done using \texttt{process.py}. Additionally, \texttt{report.py} can be used to generate metadata and report on the processed data. For simulations involving other maps, the procedure outlined in \cite{amiri2024patterns} can be applied.

\vspace{-0.2cm}

\section{Conclusion}
\label{sec:conclusion}

This paper presents a comprehensive and flexible simulated human mobility dataset designed specifically to facilitate the study of trajectory anomaly detection. By injecting various types of anomalies into an existing pattern of life simulation, including hunger, social, work, and interest anomalies, we provide a framework for researchers to develop and test anomaly detection algorithms. Our dataset addresses critical gaps in real-world data availability, particularly the lack of ground truth, by offering clear labels and varying levels of anomaly intensity. The three distinct methods of introducing anomalies—centralized assignment, infectious spread, and location-based spread—further enhance the dataset's versatility, allowing it to be used across a wide range of research scenarios. This work not only contributes a valuable resource to the field but also sets the stage for future studies aimed at refining anomaly detection techniques and applying them to increasingly complex urban mobility datasets. Future research may focus on expanding the types of anomalies and refining detection algorithms to improve accuracy and applicability in real-world scenarios.
\vspace{-0.2cm}


\bibliographystyle{ACM-Reference-Format}
\bibliography{main}

\end{document}